\documentclass{nature}
\usepackage{graphicx}
\usepackage{url}
\usepackage{subfigure}
\usepackage{amsmath, amssymb, mathrsfs}
\usepackage{mathrsfs}
\usepackage{dsfont}
\usepackage{comment,color}
\usepackage{pgfplots}
\usepackage{pgf}

\usepackage{tikz}
\usepackage{units}
\title{Optical reconstruction of dust in the region of SNR RX~J1713.7-3946 from astrometric data}

\author{R.~Leike$^{1,2}$, S.~Celli$^{3,4}$,  A.~Krone-Martins$^{5,6}$, C.~Boehm$^{7}$,  M.~Glatzle$^{1,8}$, Y.~Fukui$^9$, H.~Sano$^9$ \& G.~Rowell$^{10}$}

\makeatletter
\let\saved@includegraphics\includegraphics
\AtBeginDocument{\let\includegraphics\saved@includegraphics}
\renewenvironment*{figure}{\@float{figure}}{\end@float}
\makeatother

\begin{document}

\maketitle

\begin{affiliations}
 \item Max-Planck-Institut f\"ur Astrophysik, Karl-Schwarzschild-Str. 1, D-85741 Garching, Germany
 \item Ludwig-Maximilians Universit\"at M\"unchen, Geschwister-Scholl Platz 1, M\"unchen, Germany
 \item Dipartimento di Fisica dell'Universit\`a La Sapienza di Roma, Piazzale Aldo Moro 2, 00185, Roma, Italy
 \item Istituto Nazionale di Fisica Nucleare, Sez. di Roma, Piazzale Aldo Moro 2, 00185, Roma, Italy
 \item Donald Bren School of Information and Computer Sciences, University of California, Irvine, Irvine CA 92697, USA
 \item CENTRA/SIM, Faculdade de Ci\^encias, Universidade de Lisboa, Ed. C8, Campo Grande, 1749-016, Lisboa, Portugal
 \item School of Physics, The University of Sydney, NSW 2006 Camperdown, Sydney, Australia 
 \item Physik-Department, Technische Universit\"at M\"unchen, James-Franck-Str.~1, 85748 Garching, Germany
 \item Department of Astrophysics, Nagoya University, Furocho, Chikusa-ku, Nagoya, Aichi, 464-8602, Japan
 \item School of Physical Sciences, University of Adelaide, Adelaide, 5005, Australia
 \end{affiliations}

\begin{abstract}
The origin of the radiation observed in the region of the supernova remnant RX~J1713.7-3946, one of the brightest TeV emitters, has been debated since its discovery. The existence of atomic and molecular clouds in this object supports the idea that part of the GeV gamma rays in this region originate from proton-proton collisions. However, the observed column density of protons derived from gas observations cannot explain the whole emission. Yet there could be a fraction of protons contained in fainter structures that have note been detected so far. Here we search for faint objects in the line of sight of RX~J1713.7-3946 using the principle of light extinction and the ESA/Gaia DR2 astrometric and photometric data. We reveal and locate with precision a number of dust clouds and note that only one appears to be in the vicinity of RX~J1713.7-3946. We estimate the embedded mass to $M_{dust} = (7.0 \pm 0.6) \times 10^3 \, M_{\odot}$ which might be big enough to contain the missing protons.
Finally, using the fact that the supernova remnant is expected to be located in a dusty environment and that there appears to be only one such structure in the vicinity of RX~J1713.7-3946, we set a very precise constrain to the supernova remnant distance, at ($1.12 \pm 0.01$)~kpc.
\end{abstract}

Supernova remnants (SNRs) are believed to be among the major contributors of the Galactic cosmic-ray (CR) flux observed on Earth\cite{aharonian2013}, but conclusive evidence of this statement awaits a clear and unambiguous identification of the interaction process at the origin of their electromagnetic radiation \cite{aharonian2019}. The detection of high-energy gamma rays in spatial association with dense molecular clouds in SNRs supports the paradigm for the SNR origin of Galactic CRs, as the radiation would originate from the collisions of shock-accelerated protons with the target gas located in these environments (the so-called hadronic scenario). Since accelerated electrons can also emit TeV gamma rays in absence of collisions through synchrotron or inverse Compton processes (which is dubbed as the leptonic scenario), finding structures spatially correlated with TeV emission is key to discriminate the two scenarios. 

RX~J1713.7-3946 is a shell type SNR, the brightest emitter among this source population of TeV gamma rays\cite{hessRXJ1713} and synchrotron X-rays\cite{slane,sano2015} in our Galaxy, with spectral measurements available also in radio\cite{lanzedic} and GeV gamma rays\cite{fermiRXJ1713}. The highest energy photons recorded correspond to tens of TeV, indicating that this SNR can accelerate multi-TeV particles. However to date the origin of this radiation is uncertain. Indeed, the gamma rays detected by Fermi-LAT\cite{fermiRXJ1713} in the $1$-$300$~GeV energy range show an unusually hard spectrum, which is often interpreted in favour of a leptonic origin. However, neither the simplest (one-zone, one-particle population) hadronic nor leptonic scenarios can successfully explain the entirety of observations \cite{gabici2015}. An alternative is to assume several populations of electrons\cite{yamazaki2009gamma} or, in the hadronic case, a non uniform target density \cite{gabici2014, celli2018}.

Radio to X-ray observations indicate that RX~J1713.7-3946 is located in a clumpy molecular and atomic environment. As a result, the very hard GeV
gamma-ray spectrum may result from interaction of the shock with this clumpy gas, which is affecting the propagation of CRs in there \cite{fukui2012,inoue2012,gabici2014,celli2018}.
In such an environment, the progenitor (likely a massive star terminating in a Type II supernova) has exploded within the cavity blown by the strong wind of the precedent phase of stellar evolution, while the densest components of the surrounding molecular clumps have survived against the wind. As a result, the magnetic field is expected to be strongly amplified all around the clumps because of both field compression and magneto-hydrodynamical (MHD) instabilities developed in the shock-clump interaction. As the turbulence would prevent low energy particles from penetrating the clumps, the gamma-ray spectrum arising from the massive target would be harder than the parent proton spectrum accelerated at the forward shock\cite{celli2018}. One of the main ingredients of this scenario is the presence of a significant amount of gas and/or dust in the remnant environment. Dust detection, being complementary to HI and CO studies, is crucial to derive a complete picture of the proton density distribution in the remnant region, possibly probing
the hadronic nature of the observed radiation, despite the ambiguity of spectral signatures in the gamma-ray emission. This result could eventually be confirmed by Northern-Hemisphere neutrino experiments, as KM3NeT \cite{ambrogi,km3}, as well as by precise spectroscopical and morphological gamma-ray measurements by the Cherenkov Telescope Array \cite{cta}.

So far two different methods have been used for the identification of density structures in the environment of SNRs. One relies on the presence of molecular and atomic transition lines\cite{fukui2003,sano2010,fukui2012,sano2013,moriguchi,mopra}, and the other requires detecting bright and temporally variable filamentary structures in non-thermal X-rays\cite{uchiyama}. 
First observations performed with the NANTEN telescope in the region of SNR RX~J1713.7-3946 \cite{fukui2003} indicated the presence of a dense molecular cloud of $\sim 200$ solar masses in the north-western part of the remnant shell, spatially associated with the emission peak observed by H.E.S.S. in TeV gamma rays. Here, by 'north-western' and other directional descriptors we refer to the equatorial coordinate system, similarly to Fig.~2 of \cite{moriguchi}. 
Subsequent measurements of CO transition lines identified several other massive molecular clouds in the region, each a few parsecs across and enclosing some $\sim 100$ solar masses \cite{moriguchi}. However, it was only after including HI data that a significantly better spatial match between the TeV gamma rays and the interstellar medium (ISM) proton distribution was obtained \cite{fukui2012}, as several parts of the gamma-ray shell contain a counterpart only in HI, e.g. the south and south-western locations. 

Here we suggest that a significant part of the HI proton target component which was identified in the northern rim is embedded in an extended dust cloud of mass $(7.0\pm 0.6)\times 10^3~M_\odot$, whose discovery is reported in the present letter.
We note that an attempt to estimate the amount of dust in the SNR region was already performed\cite{dobashi}. However the lack of distance information did not allow an indisputable association with the SNR environment. As such, the observed dust was most likely dominated by contributions along the line of sight. By using combined data \cite{anders2019} of ESA/Gaia DR2 data\cite{gaia2018,gaia2018b}, 2MASS \cite{skrutskie2006two}, PAN-STARRS \cite{kaiser2002pan} and AllWISE \cite{cutri2013vizier}, we are able to accurately characterize the distance of the dust. Such a method can in principle probe both dust originated from the ejecta of the SN explosion, as well as pre-existing dust from the ISM. Our dust map shows that the presence of only one major dust cloud along the line of sight which is compatible with the current distance estimate of the SNR RX~J1713.7-3946. Under the premise that SNR RX~J1713.7-3946 is located in a dusty environment, we can accurately constrain its distance using optical data. Our results appear to be compatible with current estimates inferred from radio observations and galaxy kinematics models (through the velocity range of clouds moving in the spiral arms of the Galaxy)\cite{brand93}.

We aim to take advantage of the decrease in the luminosity of background stars induced by the presence of dust, as to unveil morphological features of the environment surrounding this interesting SNR. This method allows a three-dimensional reconstruction of extinction, hence providing a measurement of the optical absorption right at the location of the remnant. Note that previous studies\cite{dobashi} have probed the total column density of extinction, i.e. integrated along the line of sight in the direction of the SNR. Such results are strongly sensitive to both the foreground and the background of the SNR itself, especially because the SNR is located at low latitudes.

\section*{The distance to the dust cloud surrounding RX~J1713.7-3946}
\begin{figure*}[ht]
\centering
    \includegraphics{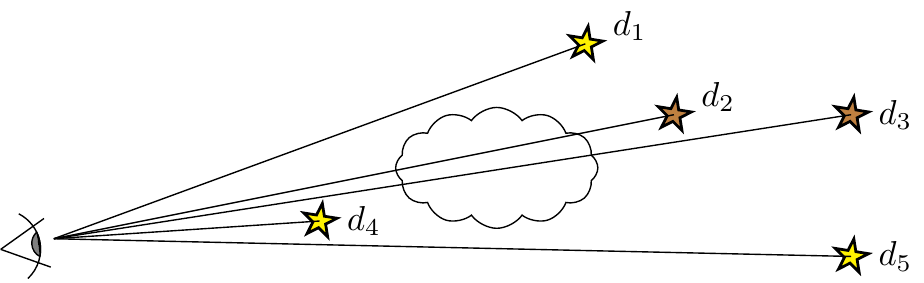}

	\caption[Measurement Principle]{\label{fig:cartoon}
	Measurement principle: starlight is reddened and extinguished when traveling through dust clouds. By comparing the extinction of many stars, the position of dust clouds can be estimated. In this figure, the dust cloud can be constrained to be located between $d_2$ and $d_4$. Note that the distances of stars are known through their parallaxes.}
\end{figure*}

Astrometric data are the most direct way to map positions, kinematics and dynamics of astronomical sources. This information is classically encoded in five astrometric parameters, namely position, parallax, and the two components of the proper motions (direction and speed), although accelerations will be available for a few sources in the near future from the ESA/Gaia space mission\cite{gaia2016mission} measurements. In principle, these astrometric data together with photometric measurements can show the presence of dust by identifying regions with strong optical extinction from background stars. Fig.~\ref{fig:cartoon} shows a schematic representation of the measurement principle.

To check for the existence of such dust components, a reconstruction needs to be performed, for instance, by considering spatial correlations of the reddening of the stars in the astrometric catalogue. In the most extreme cases, the complete lack of stellar sources in certain specific directions of the sky requires casting shadows in the background stellar distribution. In this letter, we perform a reconstruction aimed at profiling the spatial correlations of the stellar reddening.  The astrometric and photometric catalogue that we adopt to perform our analysis is obtained from a cross-matching between Gaia DR2, PanSTARRS-1, 2MASS, and AllWISE data \cite{anders2019}.
However, due to selection effects, our method is more sensitive to finding the diffuse, mildly overdense dust in the interstellar medium, than the densest clouds. This is because the chance of seeing a star behind a highly extinguishing cloud, for which the photometric measurements are exact enough to allow for the computation of a reddening estimate, are quite low.

We reconstruct the dust extinction in the SNR region using a Gaussian process. Details of the methods can be found in the Methods section or in \cite{leike2020resolving}.
For our reconstruction, we focus on the direction of the SNR, namely at Galactic coordinates $l=347.3^\circ$ and $b=-0.5^\circ$ \cite{greencat}. We select a cuboid of size 1400 pc x 128 pc x 96 pc.
The largest dimension of 1.4~kpc was chosen out of multiple considerations. On the one hand, the spatial density of sources with astrometric measurements and extinction estimates decrease with distance, an effect that renders reconstructions beyond 1.4~kpc less reliable. On the other hand, the most robust estimate of the SNR distance is $\sim 1$~kpc \cite{moriguchi}, and the SNR is extremely likely to be inside the reconstructed volume.
Our total reconstruction uses 262978 stars. Data reconstruction is performed with a voxel resolution of $0.5$~pc. Hereby the voxel resolution yields an upper bound for the angular resolution. The closest dust clouds are observed at a distance of 200~pc, where the voxel size of $0.5$~pc is equivalent to an angular resolution of $0.15^\circ$. Thus, clouds at larger distance can only be reconstructed at smaller angular resolution, since the foreground clouds can occlude them. We note also that systematic errors might occur below the $0.15^\circ$ scale. It is conceivable that a more detailed analysis could be performed when using smaller voxels, however the larger degrees of freedom render the minimization procedure inefficient as the problem already requires $\sim 100$~GB of RAM during minimization. The actual achieved resolution is of course also constrained by the number of data points per volume. This resolution can vary depending on the distance and local amount of stars. This is the main reason for adopting a method that provides us with approximate posterior samples, which enable the computation of error estimates.

\begin{figure*}[ht]
\centering
\subfigure[\label{fig:a}]{\includegraphics[trim={2.2cm 1.cm 1.2cm 2.cm}, clip, width=0.9\textwidth]{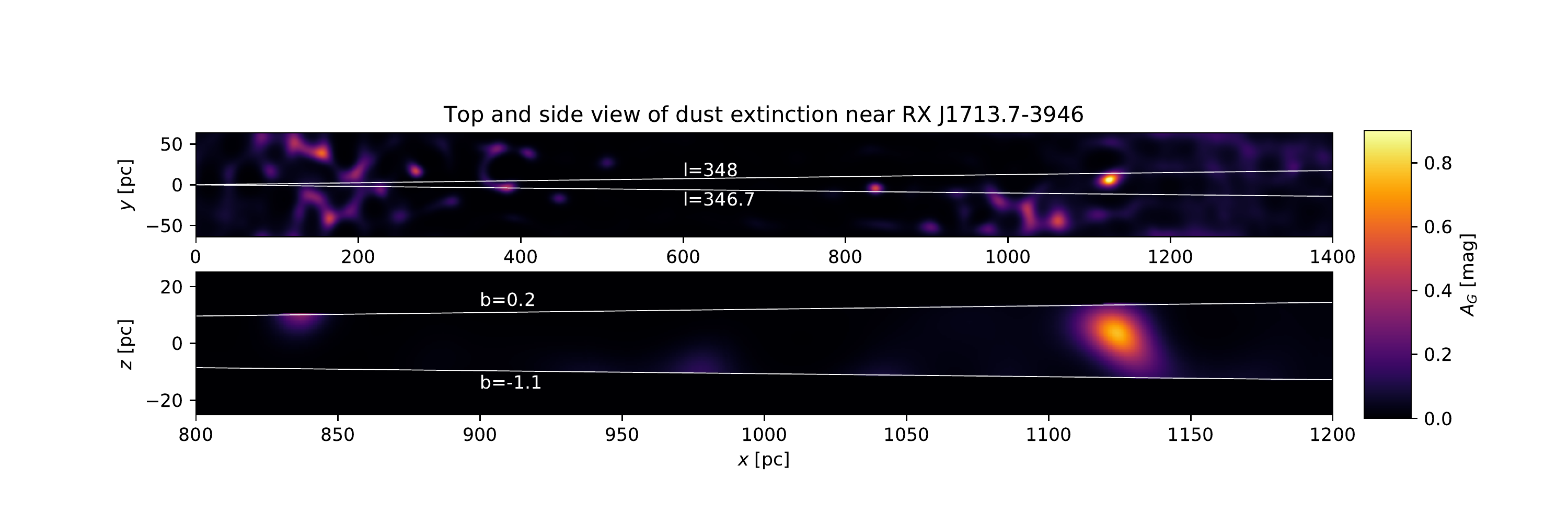}}
\vspace{.1cm}\\
\subfigure[\label{fig:b}]{\includegraphics[width=0.5\textwidth]{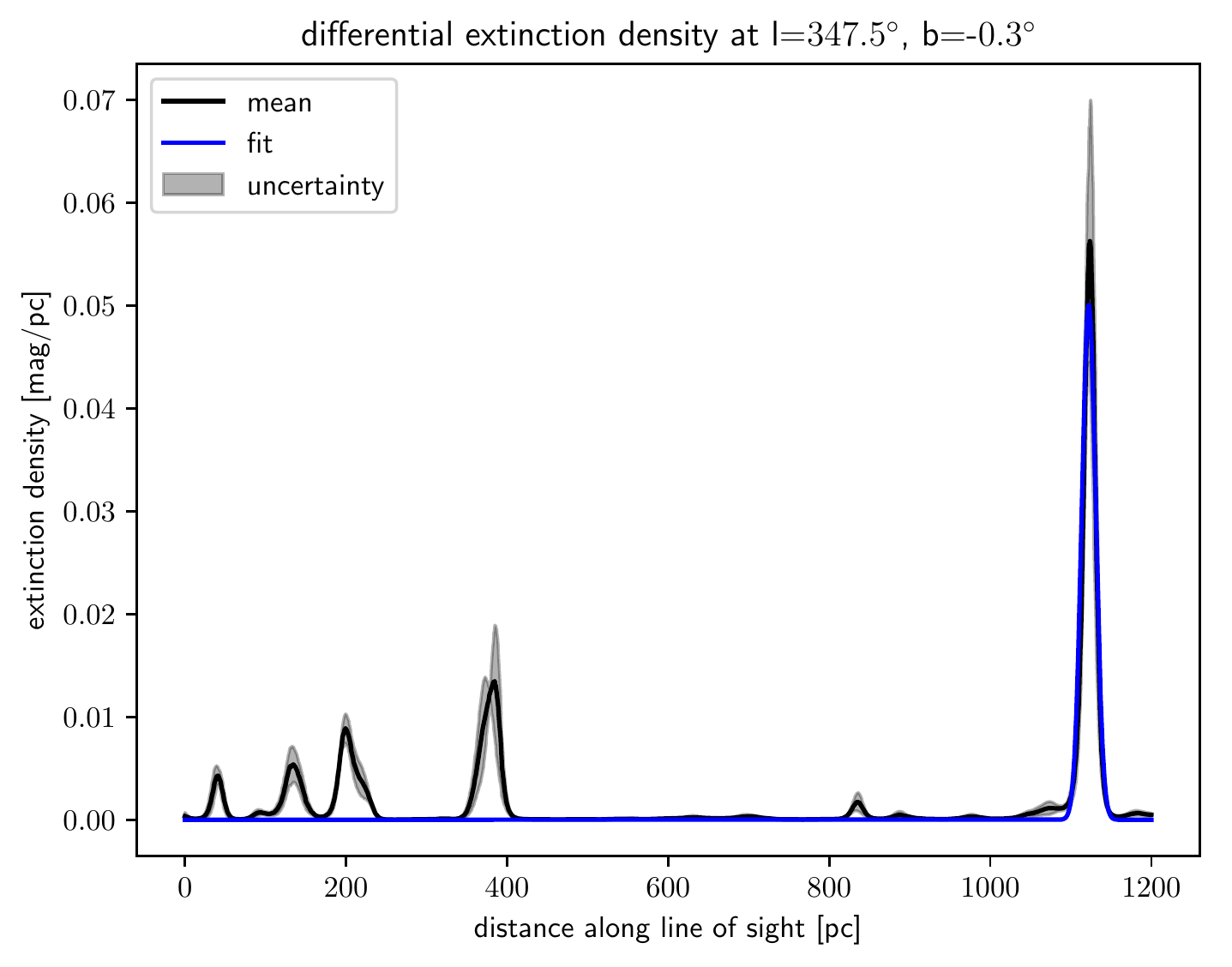}}
\caption{Reconstructed G-band extinction in the direction of the SNR. (a) Dust extinction column density in Cartesian coordinates. The Sun is at position (0,0), and the x-axis points towards the SNR ($l=347.3^\circ$, $b=-0.49^\circ$). The top panel shows the dust extinction column density within a 32~pc slice which contains the SNR. The white lines indicate longitude values of $348^\circ$ and $346^\circ$, respectively. The bottom panel views the SNR from the side, showing only extinction withs $348^\circ>l>346.7^\circ$ and $-1.1^\circ<b<0.2^\circ$. The white lines denote latitudes $b=-1.1^\circ$ and $b=0.2^\circ$, respectively.
(b) Integrated G-band extinction along the line of sight of the SNR. The gray shaded area corresponds to the $1$~$\sigma$ uncertainty region. The blue curve indicates a polynomial fit of degree two to the dust cloud around 1100~pc on logarithmic scale.}
\label{fig:dustDistance}
\end{figure*}

The amount of dust extinction observed in the G-band is shown in Fig.~\ref{fig:dustDistance}, reported in Cartesian coordinates and integrated over the vertical spatial component. This map indicates the presence of several clouds along the line of sight, most of which are closer than $400$~pc.
These foreground clouds are limiting the resolution of our present reconstruction, as the light from all stars that are further away also passes through them. As a result, our present map is not sufficiently resolved to study the detailed angular distribution of dust inside the SNR.
However, we can constrain the distance to the SNR by looking at all locations with line of sight extinction.

The strongest extinction of background light is at a distance of $(1.12 \pm 0.01)$~kpc from the Sun's location, along the Galactic Plane. At this location we report a $5.5$~$\sigma$ detection of dust, i.e. the reconstructed differential dust extinction differs from $0$ by $5.5$~$\sigma$. This is the most probable location of the SNR, under the assumption the remnant is located in a dusty environment.

We cannot exclude that there are dust clouds beyond $1350$~pc as the boundary of the reconstructed volume and beyond is hardly constrained. However, our distance reconstruction based on the distance of the highest extinction is consistent with the estimates of  $\sim 1$~kpc inferred from the measured velocity range of CO lines in the same region, ranging between $-10$~km/s and $20$~km/s \cite{fukui2003,moriguchi}.
Note that the distance estimate derived here is much more precise as it emerges from a reconstruction that combines the parallax  measurements of many stars. For instance, the immediate vicinity of the reconstructed dust cloud, i.e. between $257.75^\circ<\text{RA}<259^\circ$ and $-39.8^\circ<\text{DEC}<-39.3^\circ$ and a median distance between $900\,\text{pc}$ and $1200\,\text{pc}$, contains $578$ stars. These stars have a median parallax error of $0.1\,\text{mas}$.

\section*{Dust in the surroundings of RX~J1713.7-3946}
\begin{figure*}[ht]
\centering
\subfigure[\label{fig:a}]{\includegraphics[width=0.49\textwidth]{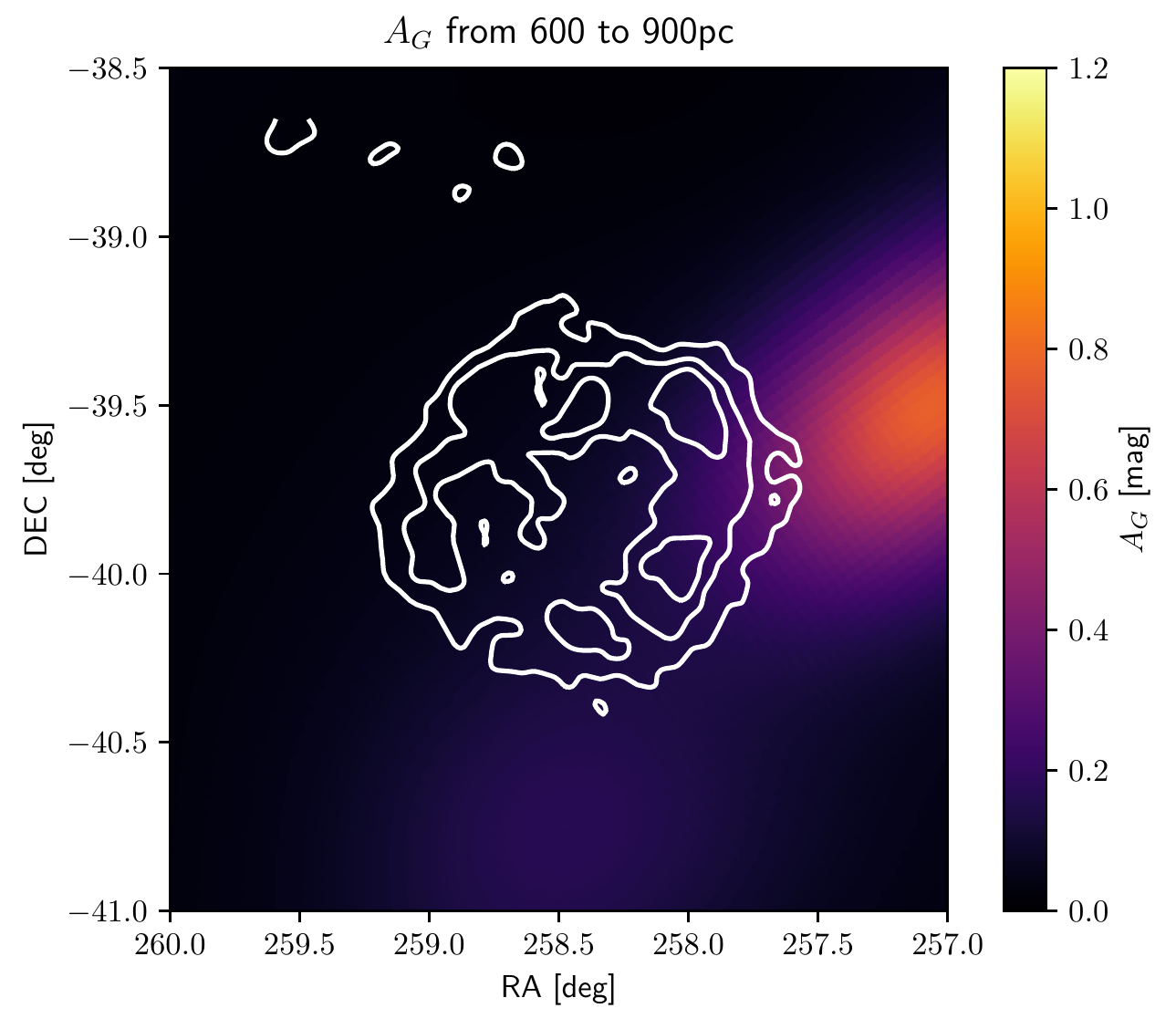}}
\subfigure[\label{fig:b}]{\includegraphics[width=0.49\textwidth]{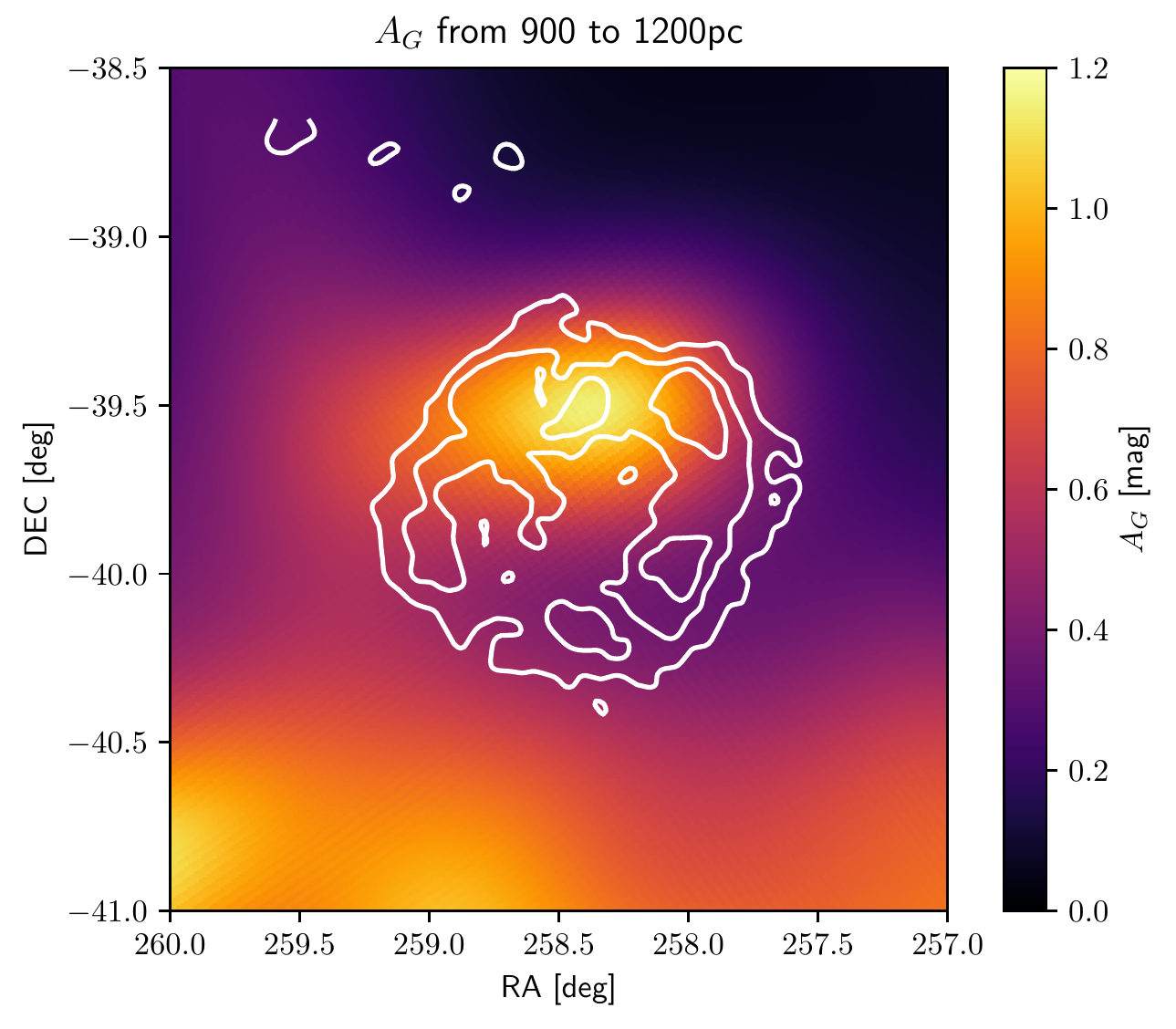}}
\caption{Reconstructed G-band extinction in the direction of the SNR, overlaid to H.E.S.S. contours of the SNR for photons with energy $>2$~TeV \cite{hessRXJ1713}. The coordinates are in the ICRS system. (a) Extinction in the 600 to 900~pc distance range; (b) Extinction in the 900 to 1200~pc distance range.}

\label{fig:dustContours}
\end{figure*}

\begin{figure*}[hb]
\centering
\subfigure[\label{fig:gasa}]{\includegraphics[width=0.45\textwidth]{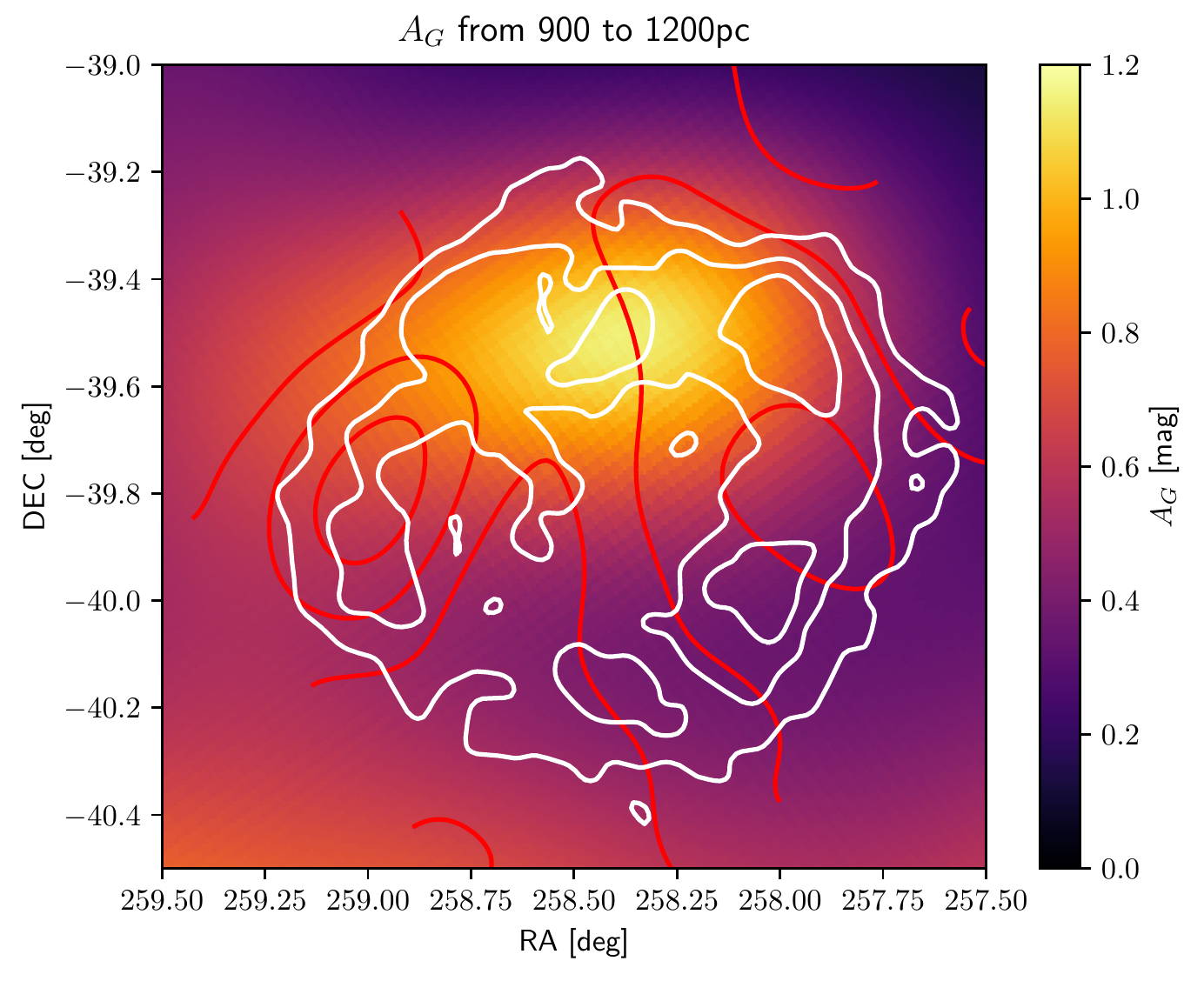}}
\subfigure[\label{fig:gasb}]{\includegraphics[width=0.45\textwidth]{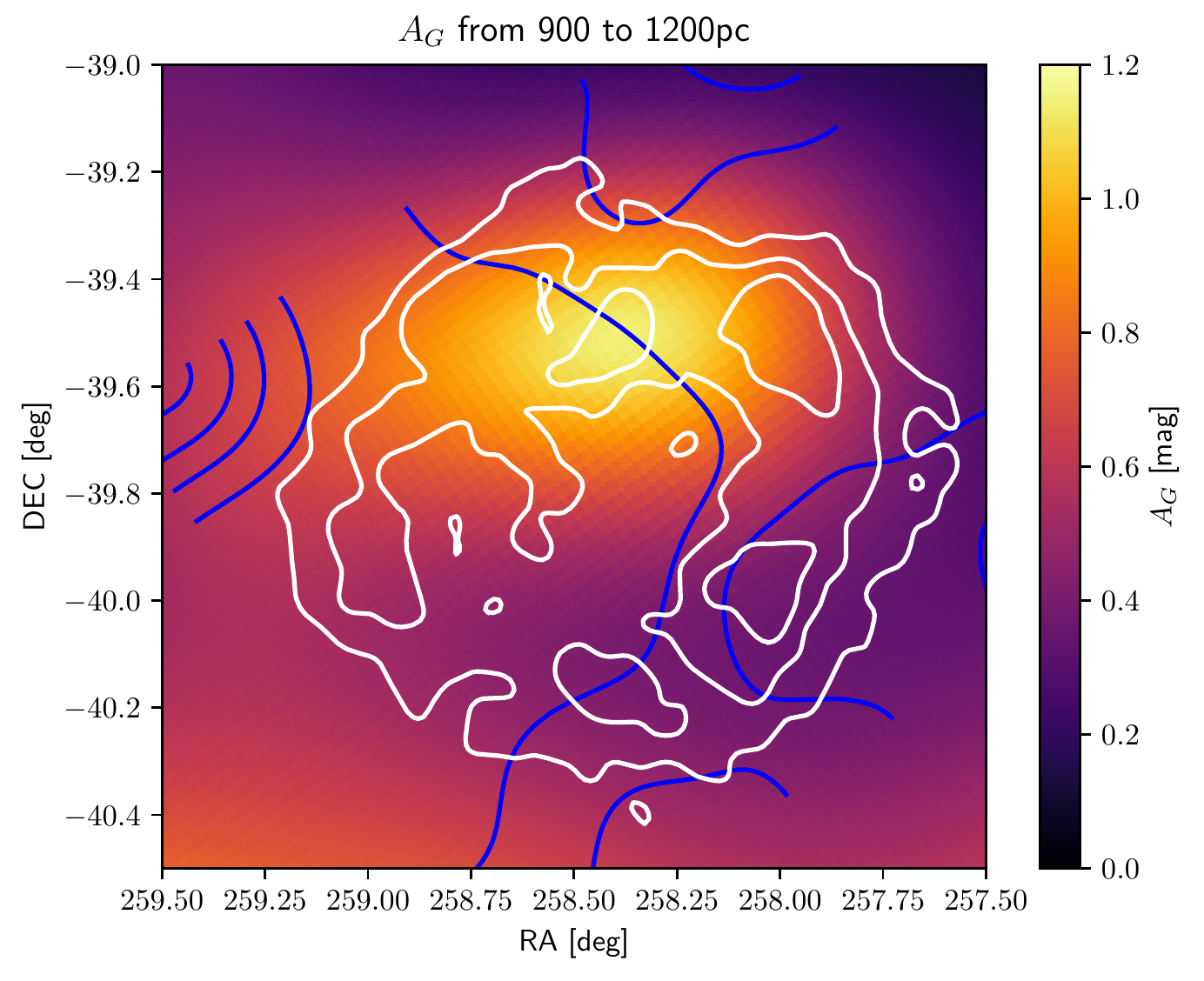}}
\subfigure[\label{fig:gasc}]{\includegraphics[width=0.52\textwidth]{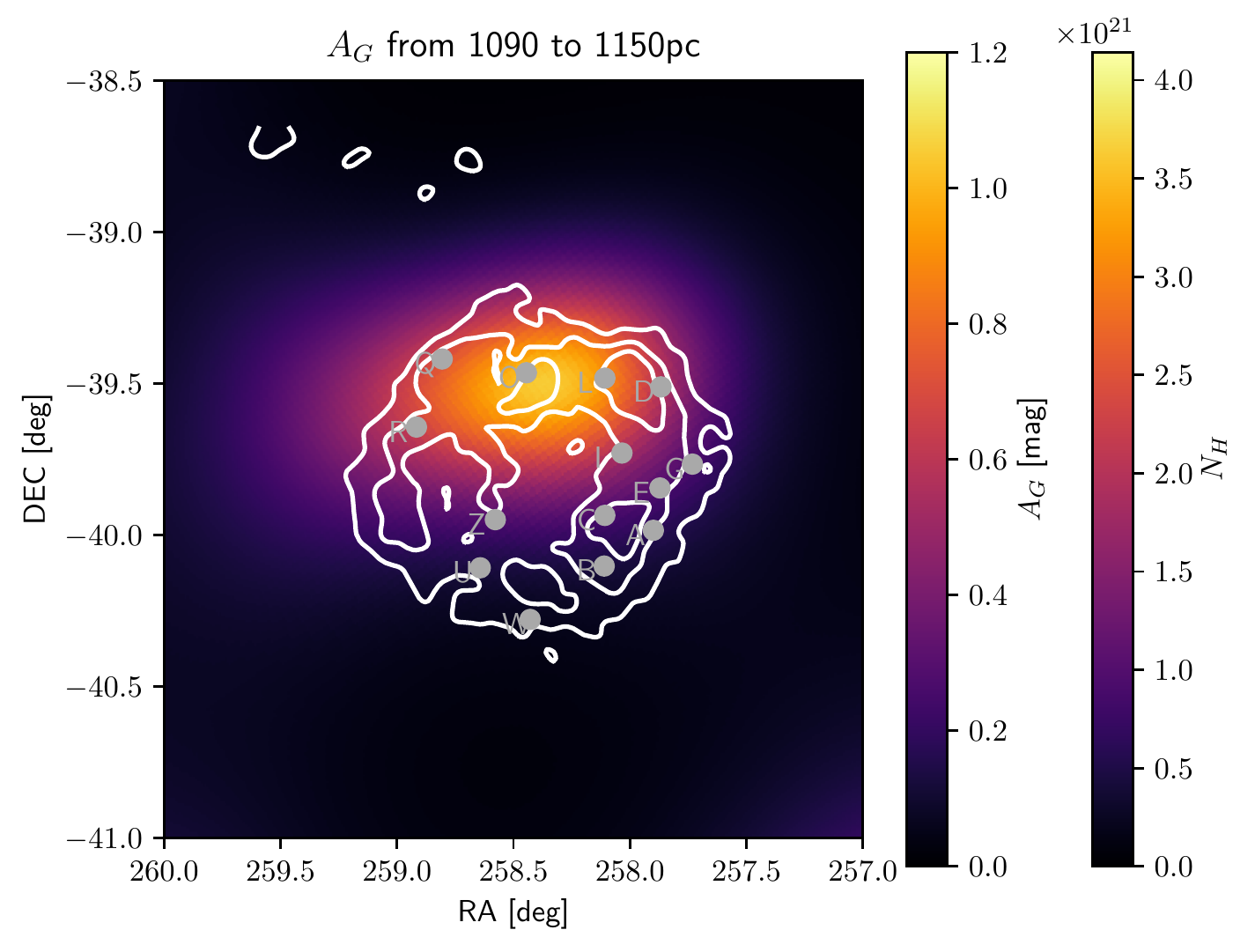}}
\subfigure[\label{fig:gasd}]{\includegraphics[width=0.45\textwidth]{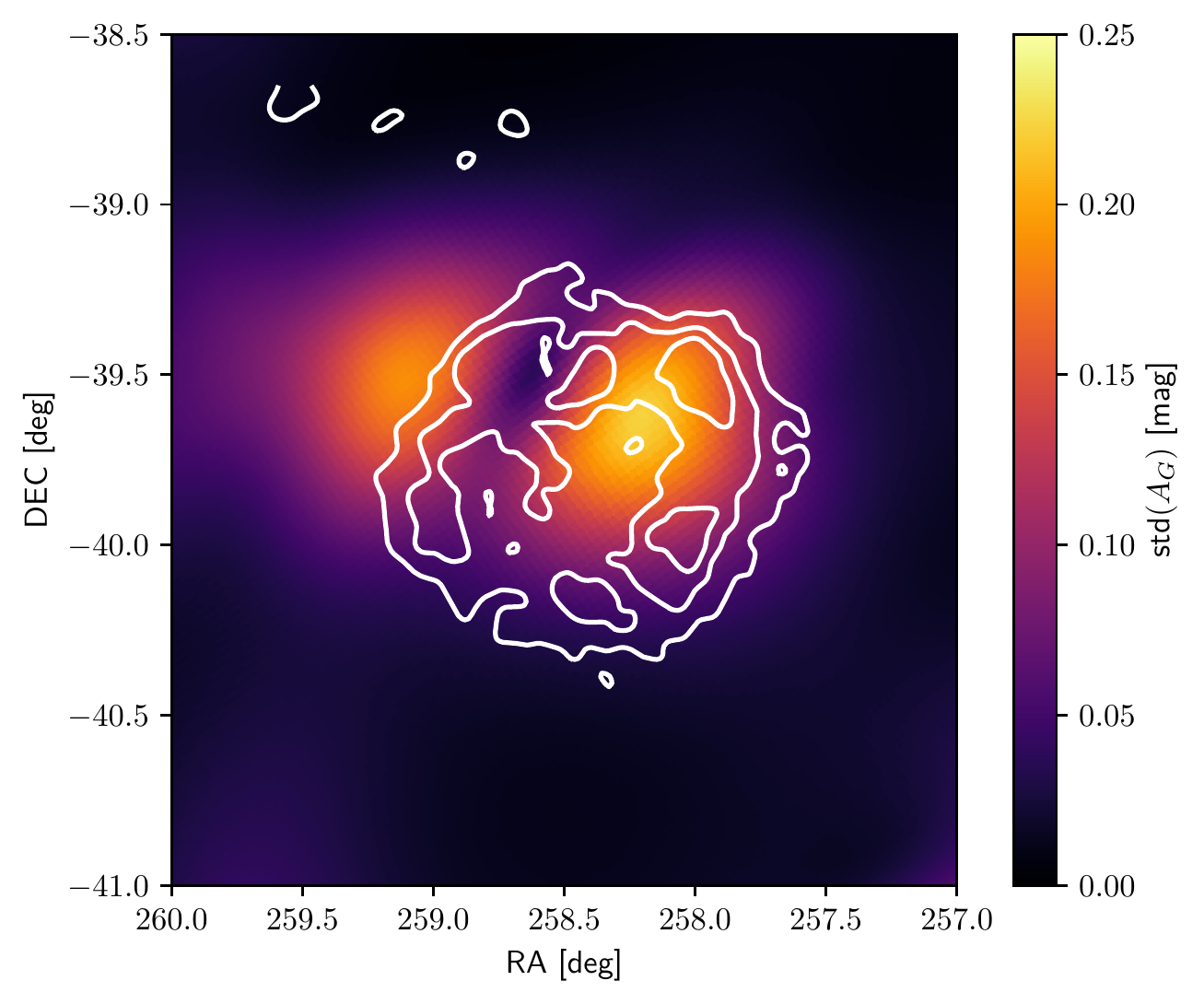}}
\caption{Reconstructed G-band extinction in the distance range: (a),(b) from 900 to 1200~pc; (c),(d) from 1090 to 1150~pc. In all panels, white contours identify the gamma-ray counts observed by H.E.S.S. above 2~TeV \cite{hessRXJ1713}. Red contours (a) refer to $N_{\rm p}$(HI) as derived using the SGPS HI data (ATCA \& Parkes), corrected for self-absorption \cite{fukui2012}, blue contours (b) to $N_{\rm p}$(H$_2$) derived using the NANTEN ${}^{12}$CO(J=1-0) data \cite{fukui2012}.
Atomic and molecular data select clouds with velocity $-20 ~\text{km/s} <v< 0 ~\text{km/s}$, corresponding to a distance estimate of $\sim 1\,\text{kpc}$. They are smoothed with $\sigma=0.1^\circ$, to match the resolutions.
In panel (c) we also show massive clouds of CO\cite{moriguchi}. Panel (d) shows the uncertainty in standard deviations.}
\label{fig:dustGasGamma}
\end{figure*}

Starting from the most likely location of the SNR, we perform a scan towards the remnant direction in order to remove the contamination from dust extinction along the line of sight, and focus only on the dust located at the remnant position. The results of such a scanning in the SNR direction are shown in Fig.~\ref{fig:dustContours} for two different distance ranges: dust located 600 to 900~pc away from the Sun is shown in Fig.~\ref{fig:a}, while dust from 900 to 1200~pc is shown in Fig.~\ref{fig:b} (both in the remnant direction).
We identify a dust cloud located inside the TeV contours of the remnant at the distance range from 900 to 1200~pc. However, at this distance we note some extinction features in the southern region of the map. By moving in the distance range from 1090 to 1150~pc, the uncorrelated feature disappears, and we are left with the dust in the remnant region. See Fig.~\ref{fig:gasc} for a plot with this distance constraint, and Fig.~\ref{fig:gasd} for the corresponding uncertainty map. We note an asymmetric distribution of the dust in the SNR region, being only located in the northern part of the remnant. 
The associated dust does not correlate with the peaks that have been observed in the very-high-energy map, where the TeV gamma rays mostly concentrate in north-western part the SNR. 
However, we note that the northern region is generally bright in gamma rays with energy above 2~TeV, and it is populated by several massive clouds\cite{moriguchi}, some of which spatially overlap with the observed dust cloud, namely core D ($M=292\, M_\odot$), core L ($M=370 M_\odot$), core O ($M=61 \, M_\odot$), core Q ($M=108 \, M_\odot$), and core R ($M=67 \, M_\odot$). 

Given this result, we estimate the column density of the dust in the SNR region to compare it with that in the gas. By adopting a linear transformation between G-band extinction and column density of protons in the ISM (as described in the Methods), we obtain the results shown in Fig.~\ref{fig:gasc}. This result refers to a distance range from $1090$ to $1150$~pc. The figure shows that the amount of proton column density of the dust cloud is at most $N_{\rm p} = 4 \times 10^{21}$~cm$^{-2}$. This value has to be compared with a column density in total gas (HI+H$_2$) ranging between $5 \times 10^{21}$~cm$^{-2}$ (towards the south-east of the remnant) and $13 \times 10^{21}$~cm$^{-2}$ (towards its south-west). The gas column density values have been estimated by accounting for self-absorption in the HI component\cite{fukui2012}. 
A visual comparison to the HI and H2 components is displayed in Figs.~\ref{fig:gasa} and ~\ref{fig:gasb}, where we overplot our found dust extinction with ATCA data and NANTEN data, respectively.
Hence, the dust cloud encloses a minor fraction of target protons with respect to all of the gas clouds located in the remnant environment. 
We remark that the protons embedded in the dust do not necessarily represent an additional component with respect to previously determined gas level in the northern region. However, if this is indeed the case, it would yield an absorption column density in the northern region matching the one derived from X-ray mapping, up to $N_{\rm p} \simeq 1 \times  10^{22}$~cm$^{-2}$ (see Fig.~3 in \cite{sano2015}). As such, this dust cloud has the potential to trace additional material, possibly acting as a target for proton-proton collisions in the northern region.

The newly detected dust cloud encloses a total mass of $M_\text{dust}=(7.0 \pm 0.6)\times10^3M_\odot$. Such a high mass value is incompatible with a SN ejecta origin since ejecta are  expected to be considerably less massive than the above estimate \cite{micelotta}. Hence, the only possibility is that the dust cloud is a pre-existing component of the ISM. This interpretation is also consistent with the observed morphology of the dust feature, whose asymmetric distribution in space with respect to the SNR is most likely connected to the parent molecular cloud.

We also highlight, that using the Herschel, Spitzer and Wise data, we found no evidence of infra red emission at the location of this faint structure. While this warrants the need for further investigations, we note that a mixture of carbon and silicate might explain this high level of IR extinction, especially since silicate dust formed in the atmospheres of late-type stars have already been proposed to explain the extinction in the inner kpc of the galaxy \cite{ Voshchinnikov_2017}.  In any case this feature should help in constraining the nature of the dust grains contained in this structure and specific SNR environment \cite{ Draine:2003if,Xue_2016, Wang_2019}.

\section*{Conclusions}
RX~J1713.7-3946 is among the most studied SNRs at very high energies, thanks to its strong X-ray and TeV gamma-ray fluxes. Despite the numerous studies, so far the distance to this source has been poorly constrained, though this is a crucial parameter to understand the evolution of the SNR, and infer e.g. the energy released at the SN explosion, the shock speed, and the remnant age. The superior quality of ESA/Gaia astrometric data in the region of SNR RX~J1713.7-3946 reveals an extended dust cloud at the remnant position, thus offering a unique opportunity to build a precise map of the proton distribution surrounding the remnant. The cloud location partly overlaps the TeV gamma-ray emission as observed by H.E.S.S. \cite{hessRXJ1713}, being mostly concentrated in the northern part of the SNR shell. The estimated embedded mass of $M_\text{dust}=(7.0 \pm 0.6)\times10^3M_\odot$ points towards the interpretation that the dust feature originates mostly from a pre-existing cloud of the ISM. However, as this is the only feature detected at the estimated distance of the SNR, it might possibly enclose also part of the SN ejecta. Hence, the presence of such dust cloud provides a direct method for estimating the distance of the SNR, which more precise than other methods such as the detection of rotational lines in the radio domain, and their association to the rotation curve of the Galaxy. As a result of this study, the distance value obtained is $d=(1120 \pm 10)$~pc.

We note, however, that at the level of our reconstruction based on current ESA/Gaia data, we are strongly limited by angular resolution and parallax uncertainty. With the upcoming data release, we expect the number of usable stars to increase and their parallax uncertainty to decrease. Moreover, the better constrained parallax uncertainties will lead to better constrained reddening estimates. All of these affect the quality of our three dimensional dust reconstruction. Thus, the spatial reconstruction that we obtain here is expected to be significantly improved with the forthcoming Gaia data releases.





\begin{addendum}
\item[Correspondence] Correspondence and requests for materials should be addressed to R. Leike (email:reimar@mpa-329garching.mpg.de), S. Celli (email: silvia.celli@roma1.infn.it).
\item[Acknowledgements] This research has made usage of public data from the H.E.S.S. telescope \url{https://www.mpi-hd.mpg.de/hfm/HESS/pages/dl3-dr1/}. The authors gratefully acknowledge the NANTEN team, for sharing gas data in our region of interest. AKM acknowledges the support from the {\it Funda\c c\~ao para a Ci\^encia e a Tecnologia} (FCT) through grants PTDC/FIS-AST/31546/2017 and UID/FIS/00099/2013. This work has made use of data from the European Space Agency (ESA) mission {\it Gaia} (\url{https://www.cosmos.esa.int/gaia}), processed by the {\it Gaia}
Data Processing and Analysis Consortium (DPAC,
\url{https://www.cosmos.esa.int/web/gaia/dpac/consortium}). Funding for the DPAC
has been provided by national institutions, in particular the institutions
participating in the {\it Gaia} Multilateral Agreement. Some of the authors are members of the {\it Gaia} Data Processing and Analysis Consortium (DPAC).

\item[Author Contributions] The idea was proposed by C.~Boehm and developed with S.~Celli, A.~Krone-Martins. R.~Leike has developed and applied the entire method, written the method section and contributed to the text of the manuscript.
 S.~Celli has written large parts of the manuscript and helped reviewing and understanding results of the method.
 A.~Krone-Martins and C.~Boehm have written significant parts of the manuscript and helped reviewing and understanding results of the method.
 M.~Glatzle has contributed code and numerical advice regarding the method and helped in reviewing the manuscript.
 Y. Fukui and H. Sano have contributed ISM data and helped in reviewing the manuscript. 
 G.~Rowell has contributed to the data and helped in reviewing the manuscript and understanding the results. 
\item[Competing Interests] The authors declare that they have no competing financial interests.f

\end{addendum}
\newpage
\listoffigures

\begin{methods}
\label{sec:methods}
We infer the dust extinction density using a Bayesian approach. 
The reconstruction is performed in Cartesian coordinates.
Our reconstructed volume has a size of 1400 pc x 128 pc x 96 pc, with the x-axis pointed towards $l=347.3^\circ$, $b=-0.49^\circ$, the y-axis towards $l=77^\circ$, $b=-0.11^\circ$, and the z-axis is chosen perpendicular to the other axes.

There are three main ingredients that enter the inference: i) the likelihood, ii) the prior, and iii) the inference method used. Before discussing each item individually, we comment on the data selection process. Lastly, we discuss the post-processing performed on reconstructed data, to extract the proton density of the dust as well as the distance estimate to SNR RX~J1713.7-3946.

\subsection{Data selection}

Among the data set defined in \cite{anders2019}, we select all stars that are within the reconstruction volume according to their $84\%$
distance quantile. We also select them for quality, namely we require a clean Gaia flag and Starhorse flag and a parallax error lower than $30\%$. Furthermore we do not use stars for which the $5\% A_V$ quantile is equal to the $16\% A_V$ quantile, as that suggest that there were problems with the sampling of this data point.

\subsection{Likelihood for STARHORSE data.}

The likelihood $P(d\vert\rho)$
states how likely the extinction data $d_A$ are given the unknown distribution of dust $\rho$.
We use data from the starhorse pipeline \cite{anders2019} obtained by cross-matching Gaia DR2 with the PanSTARRS-1, 2MASS, and AllWISE catalogues.
This data set\cite{anders2019} only contains posterior dust extinction quantiles in the V-band, from which there is no natural way to build a likelihood.
In order to build a likelihood, we use the median G-band extinction of the starhorse data situated in regions, where the Planck dust map\cite{akrami2018planck} shows minimal amounts of dust ($<e^2\frac{\mu K}{RJ}$).
We compute the mean $m_0$ and standard deviation $\sigma_0$ for the G-band extinction $d_A^{(i)}$ of all stars in these regions:
\begin{align}
m_0 &= \frac{1}{N}\sum_{i=1}^{N} d_A^{(i)}\\
\sigma_0 &= \sqrt{\frac{1}{N(N-1)}\sum_{i=1}^{N} (d_A^{(i)}-m_0)^2}
\end{align}
where the index $i$ runs over the various stars in the catalog. We found $m_0=0.1261\,\text{mag}$ and $\sigma_0=0.2421\,\text{mag}$, which we round to $\sigma_0^\prime=0.25\,\text{mag}$. This yields the likelihood
\begin{align}
P(d_A^{(i)}|A_i) = \mathscr{G}(d_A^{(i)}-m_0-A_i, \left(\sigma_0^\prime\right)^2)
\end{align}
where $A_i$ is the true extinction obtained by integrating the true density $\rho$ over the line of sight towards the source $i$, and taking the expectation value with respect to the unknown parallax.

\subsection{Correlation prior.}

We assume the dust extinction density $\rho$ to be distributed according to a Log-Normal process:
\begin{align}
\rho = \mathrm{exp}(s)\ ,
\end{align}
with $s$ an {\it a priori} isotropic and statistically homogeneous Gaussian random field. This implies that the 2-point correlation function of $s$ is diagonal in Fourier space and can be characterized by a power spectrum $p_s(|k|)$
\begin{align}
\left<ss^\dagger\right>_{P(s)} = S = \mathbb{F}^{-1}p_s(k)\mathbb{F}^{-1\dagger}\ ,
\end{align}
where we denote with $\mathbb{F}$ the Fourier transform.
We infer the power spectrum $p_s(|k|)$ non-parametrically as well, assuming it is differentiable and has a tendency to fall.
The overall latent parameters of the models are $\xi=(\xi_0, \xi_1)$, which are related to the dust extinction density $\rho$ as
\begin{align}
\rho = \mathrm{exp}(\mathbb{F}^{-1}\left(\xi_1(k)p_s(\xi_2)(|k|))\right)\ .
\end{align}
Our hyperparameter setting and prior model was already successfully applied for the inference of nearby dust clouds \cite{leike2019charting} and \cite{leike2020resolving}, where the exact same prior model was used.

\subsection{Variational inference.}

In order to derive meaningful posterior statistics, we use metric Gaussian variational inference\cite{knollmller2019metric}.
In this approach, the posterior is approximated with a Gaussian distribution $\mathscr{G}(\xi-m,D)$, where $m$ is the inferred latent mean parametrizing the dust density $\rho$ and its power spectrum $p_s$, and $D$ is the approximate posterior covariance, which is
identified with the inverse Bayesian Fisher Metric at the inferred mean:
\begin{align}
D = \left(\left<\frac{\partial H(d\vert m)}{\partial m}\frac{H(d\vert m)}{\partial m^\dagger}\right>_{P(d|m)}+
\left<\frac{\partial H(\xi)}{\partial \xi}\frac{H(\xi)}{\partial \xi^\dagger}\right>_{\delta(\xi-m)}    \right)^{-1}
\end{align}
where $H$ denotes the negative logarithm of probability.
The latent mean $m$ is obtained by minimizing the Kullback-Leibler divergence between the approximate and the true posterior. The MGVI algorithm allows for a variational inference with full covariance at similar computational cost to the maximum-a-posteriori approach\cite{knollmller2019metric}.
The output of the method is a set of approximate posterior samples.
We use these posterior samples to calculate the expectation value and uncertainty estimates of the dust extinction density $\rho$, and any of our derived quantities.

\subsection{Postprocessing.}

Using our G-band extinction density, we calculate the proton density of the ISM.
For that we invoke the fitting relation\cite{foight2016probing}
\begin{align}
N_{\rm p} = (2.87\pm 0.12)\times10^{21}A_V\,\text{cm}^{-2}\,\text{mag}^{-1}\ .
\end{align}
We supplement this correlation by a relation of G-band and V-band extinction inside the starhorse data\cite{anders2019} :
\begin{align}
A_V = 1.202\pm 0.047  A_G
\end{align}
Adding the relative errors in quadrature, we get to the final relation of
\begin{align}
N_{\rm p} = (3.45\pm0.20)\times10^{21} A_G\,\text{cm}^{-2}\,\text{mag}^{-1}\ ,
\end{align}
i.e. a total relative error of $6\%$. Note that this ratio might be subject to systematic errors if the gas to dust ratio is different within the SNR

Using this relation we can also compute the total mass of the found dust cloud at 1.12~kpc. We do so by integrating all dust located between $346 ^\circ< l < 348^\circ$, $-1.5^\circ < b< 0.5^\circ$ and $1090\,\text{pc}<r<1150$~pc, and find the total mass to be
\begin{align}
    M_\text{dust}=(7.0 \pm 0.6)\times10^3M_\odot,
\end{align}
accounting for uncertainties in the dust map and the relative error derived directly above.

Furthermore, we obtain the distance estimate to the cloud by using the line of sight extinction density displayed in Fig.~\ref{fig:dustDistance}~(b).
We fit a second order polynomial to every posterior sample of the logarithmic differential extinction density at distances from $1110$ to $1150$~pc.
The fit results in a peak distance of ($1124\pm 2$)~pc and a width of ($10.4\pm 1.0$)~pc along the line of sight.
The uncertainty of $2$~pc is obtained by the varying cloud distance in different posterior samples, while the width of $10.4$~pc results from both the smearing induced by the reconstructed correlation as well as from the physical extension of the cloud. 
For our reported distance estimate, we regard the fitted width as independent uncertainty and thus add the $10.4$~pc to the $2$~pc in squares to arrive at the final distance estimate of ($1124\pm 11$)~pc.

\subsection{Comparison to CO data.}

To validate our results and to illustrate the achieved resolution, we compare our findings to data of the NANTEN telescope. The correspondence between dust and CO is not expected to be perfect since they could also potentially represent different constituents of the ISM \cite{Goodman:2008hq}. Additionally, CO lines provide only an indirect measurement of distances through their Doppler shift, which can only be used to determine a distance when assuming a rotation curve for the Milky Way.
The comparison is furthermore complicated due to the low angular resolution of our dust reconstruction.
Nonetheless, we find that the foreground cloud with highest extinction between $350\text{-}400\,\text{pc}$ has a CO-counterpart in velocity ranges from $5\text{-}10$~km/s, as can be seen in Fig.\,\ref{fig:near-cloud-comparison}. 

\end{methods}

 \section*{Data Availability}
 The data used for our reconstruction is available under the DOI \textbf{10.1051/0004-6361/201935765}.
 The result of our reconstruction is made available under the DOI \textbf{10.5281/zenodo.4462826}. It consists of 6 approximate posterior samples over the whole reconstruction volume in Cartesian coordinates.
\newpage

\begin{figure}
    \centering
    \includegraphics{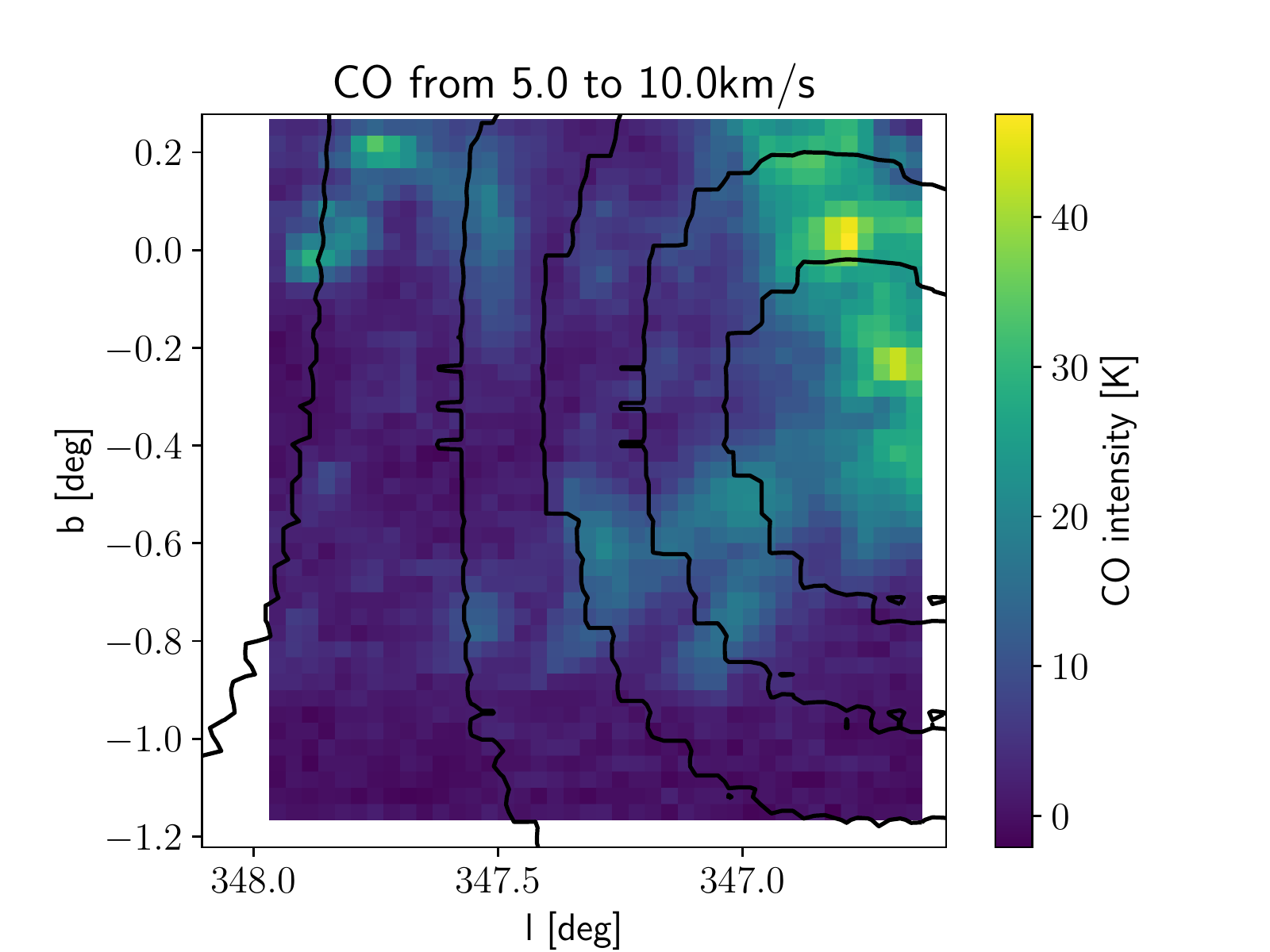}
    \caption{Comparison between CO clouds in the velocity range from $5$ to $10$~km/s,
    and the largest reconstructed foreground dust cloud (black contours) located between $350$  to $400\,\text{pc}$.
    The black contours show $0.15$ increments of G-band extinction. The resolution of our dust reconstruction has vastly inferior angular resolution and captures the morphology seen in CO on large scales of about $0.3~\text{deg}$.}
    \label{fig:near-cloud-comparison}
\end{figure}

\end{document}